\documentclass{elsart}

\usepackage{graphicx,amssymb,amsmath}
\usepackage{bm}

\begin{document}

\begin{frontmatter}

\title{Predictions of striking energy and angular dependence in
$pp\rightarrow (pp)_{S{\rm -wave}}\pi^0$ production}
\author{J.A. Niskanen}

%\affiliation{
\address{
Department of Physical Sciences, PO Box 64, FIN-00014 University of
Helsinki, Finland }

%\date{\today}

\begin{abstract}
A phenomenological calculation from threshold to 800 MeV of the
initial proton beam energy is presented to describe recent data on
the reaction $pp\rightarrow (pp)_{S{\rm -wave}}\pi^0$ with a low
energy cut on the final state diproton excitation energy. A strong
forward dip is obtained in the differential cross section as in the
data from COSY at 800 MeV, although the absolute value of the
forward cross section is too low. Earlier low energy data from
CELSIUS are reasonably well reproduced. In the unexplored energy
interval between these two experiments the model predicts a
spectacular energy dependence both in the forward direction and in
the angle-integrated cross section.

\end{abstract}

%\\ Keywords:  one; two; three}

\end{frontmatter}
%\maketitle
\section{Introduction}
Pion production cross sections in two-nucleon collisions have in a
broad sense existed for a long time (for a historical reference see
Ref. \cite{Lock}, a modern review close to threshold Ref.
\cite{hanhart}). However, only quite recently have experiments on
$NN \rightarrow NN\pi$ reactions with a cut-off on the final $NN$
excitation energy opened a new chapter in comparison with theory.
Restriction to only one $NN$ partial wave ($S$ wave) simplifies the
comparison tremendously to be basically similar to the simple
$NN\rightarrow d\pi$. It is clear that in this kind of experiments
good resolution of momenta is essential and cooled beams give an
obvious advantage, although such experiments were initiated at
TRIUMF with measurements of the differential cross section and
analyzing power $A_y$ in quasifree $pn\rightarrow (pp)_{S{\rm
-wave}}\pi^-$ \cite{Ponting,Duncan,Hahn}. In these experiments a
cut-off of $\approx 1.5$ MeV was applied on the final diproton
energy (37.5 MeV/c on the canonical c.m. momentum \cite{davepat}).
The data agreed reasonably well with the predictions of Refs.
\cite{piminus,impact} for the inverse quasifree absorption of
negative pions on the $^1S_0$ $pp$ pair in $^3$He.

Later differential cross sections between 310 and 425 MeV for
$pp\rightarrow (pp)\pi^0$ have been obtained at CELSIUS both
integrated over the final momentum magnitudes (and the nucleon
relative angle) and also applying a diproton energy cut of 3 MeV (53
MeV/c momentum) \cite{Bilger}. In the latter case the final diproton
should be rather purely $S$ wave. An interesting feature was  that
above 350 MeV the slope of the angular distribution applying the
energy cut was opposite as compared with the case without the cut.
Normally the cross sections tend to find a maximum in the forward
direction. However, the  $pp\rightarrow (pp)_{S{\rm -wave}}\pi^0$
cross section decreases for the decreasing reaction angle. This is
in agreement with the predictions given already in Ref.
\cite{piminus} for pion absorption on a $pp$ pair in the
corresponding isospin situation. A similar behaviour is also seen in
a very recent measurement by the ANKE collaboration at COSY of this
reaction at 800 MeV very near the forward direction \cite{Dymov}.

\section{Model}
The basically phenomenological model has been presented in the past
in some detail for mechanisms in Refs. \cite{piminus,impact} (albeit
for pion absorption on a bound diproton) and for the treatment of
the long range free nucleon wave functions and the Coulomb
interaction in Ref. \cite{endep}. The mechanisms in the production
operator involve first the direct production from each nucleon with
distorted initial and final $pp$ states. This is Galilean invariant
with the axial current part $\propto {\bf q}\cdot \bm{\sigma}$ and
the corresponding recoil term (axial charge) $\omega_{\bf q}{\bf
(p+p')}\cdot\bm{\sigma}/2M$. In pion reactions the all important
$\Delta(1232)$ resonance is treated as excitation of a $\Delta N$
intermediate state by coupled channels with a transition potential
including pion and $\rho$ meson exchange. This covers pion
rescattering in the $\pi N \; p_{3/2}$ partial wave. Following Ref.
\cite{impact} pion $s$-wave rescattering from the second nucleon is
parameterized as occurring on the energy shell with the
corresponding propagator for a pion emitted by a $\Delta$ taken also
on shell and off shell as in e.g. Ref. \cite{Koltun} if emitted by a
nucleon \footnote{On chiral perturbation arguments Ref.
\cite{Lensky} suggests putting the pion on shell also in this
case.}. A monopole form factor with the cut-off mass  550 MeV  is
also included in the exchange. Further, the heavy meson exchange
effect suggested in Refs. \cite{Lee,Horowitz} to account for the
missing $pp\rightarrow (pp)\pi^0$ threshold strength \cite{Meyer} is
used \footnote{As another possibility to account for this at least
partially off-shell pion rescattering has been proposed
\cite{Oset,HH}.}. The latter is implemented as in Ref. \cite{impact}
fitting the data at a single energy 290 MeV and the agreement with
data is very good up to $\eta = q_{\rm max}/m_\pi \approx 0.6$, i.e.
in the range where the $Ss$-wave production is by far
dominant~\cite{impact}.

The interactions thus fixed will be applied at higher energies with
a constraint of the two final state protons being in the $^1S_0$
state. Having only one $NN$ final state simplifies the theoretical
treatment significantly to resemble the reaction $pp\rightarrow
d\pi^+$, although the long-range behaviour of the free protons
requires some extra care \cite{endep}. When the laboratory
energy increases above, say, 350 MeV, the
final state nucleons will not remain in the $S$ wave, if the whole
phase space is included, and the results of the present calculation
involving only that should fall below any such experiment  as seen
e.g. in Ref. \cite{impact}. With a cut on the final diproton
excitation energy the validity range is increased and only the
limitations of the model itself will eventually make it fail. In
experiments this cut is the way to single out $S$-wave final nucleons.

One specific but relevant feature concerning the role of the
$\Delta$ should still be mentioned. The reactions $pp\rightarrow
d\pi^+$ and $pp\rightarrow np\pi^+$ are dominated by the $\Delta$
causing a wide peak at the nominal mass of the $\Delta$ excitation
around 600 MeV (lab). However, in the present case with a final
$^1S_0$ nucleon pair the initial state $^1D_2$ coupled to an
$S$-wave $\Delta N$ is not possible, but the $\Delta$ is excited at
least in a $p$ wave. Because of the centrifugal energy the $\Delta$
excitation should be somewhat suppressed and its position displaced
by about 80 MeV in the c.m.s. to appear most prominently at about
800 MeV laboratory energy \cite{dibaryon}. This is just the energy
of the recent COSY/ANKE experiment \cite{Dymov}. However, actually
no peaking is is seen at this energy but rather close to the nominal
$\Delta$ mass of  $\sqrt s \approx 1310$~MeV.

\begin{figure}[tb]
\begin{center}
\includegraphics[height=10 cm,angle=90]{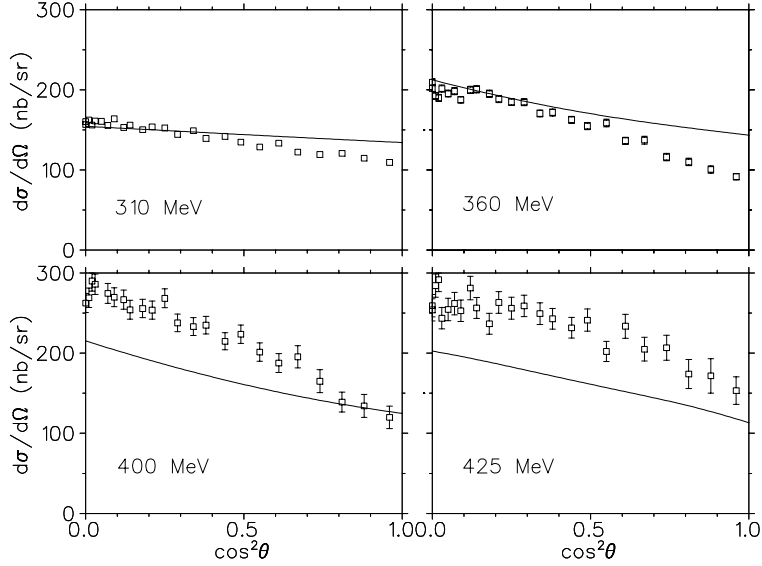}
\caption{The model vs. the Celsius data \cite{Bilger} at various
energies with a cut of 3 MeV in the relative final $pp$ energy to
constrain them to the $S$ wave.} \label{fig:bilger}
\end{center}
\end{figure}

\section{Results}

Figure \ref{fig:bilger} shows the calculation compared with a
representative selection (four of six energies with the best error
limits) of the Celsius differential cross section data
\cite{Bilger}, which are constrained to the final $^1S_0\; pp$ state
by a cut off  $E_{pp}\leq 3$ MeV on the diproton energy. The trends
are very similar and also the agreement in the absolute magnitude is
good in this energy range, although the highest maximum final pion
momentum is even twice that of Ref. \cite{Meyer}. In particular the
cross section distinctly gets a minimum in the forward direction.
This is in contrast to most other situations, e.g. the same cross
sections without the cut \cite{Bilger}.

\begin{figure}[tb]
\begin{center}
\includegraphics[width=10 cm]{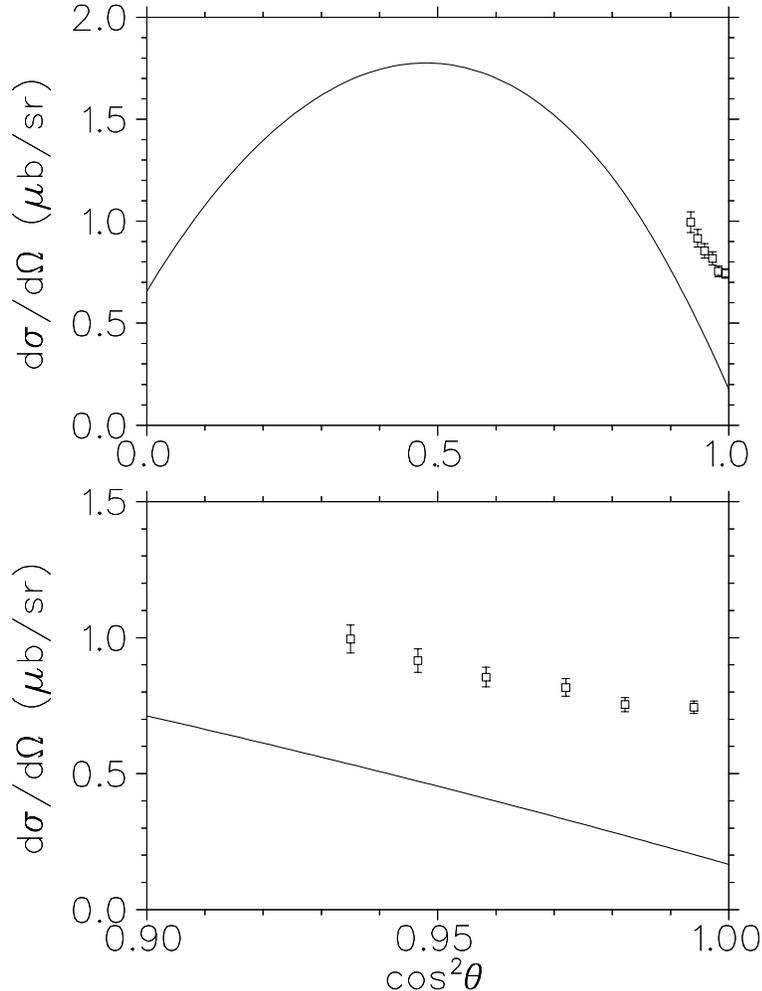}
\caption{The model results at 800 MeV. The upper panel shows the
whole angular range, while the lower shows the forward range
relevant to the COSY/ANKE experiment \cite{Dymov}, from which
the data are taken.} \label{fig:800mev}
\end{center}
\end{figure}

In the recent data \cite{Dymov} from COSY measured at 800 MeV in the
near-forward direction it was found an extraordinarily strong
angular dependence with the cross section dropping down by 30\% in
the interval where $\cos\theta$ changed only from 0.97 to 1. This
steep dip is rather unexpected even in the light of the previous
CELSIUS results showing the minimum in the forward direction. As
seen in Fig. \ref{fig:800mev} such a very steep descent is also
obtained by the present model, although the absolute scale is too
low. However, since the cross section drops by an order of
magnitude, it is clear that there is extremely strong cancellation
in the forward direction between different amplitude components
(three important ones expressed earlier in the Introduction) and so
a relatively minor change in a single partial wave may cause a large
change in the cross section. Also another minimum is predicted at
90$^\circ$ and a maximum at about 50$^\circ$.

\begin{figure}[tb]
\begin{center}
\includegraphics[height=10 cm,angle=90]{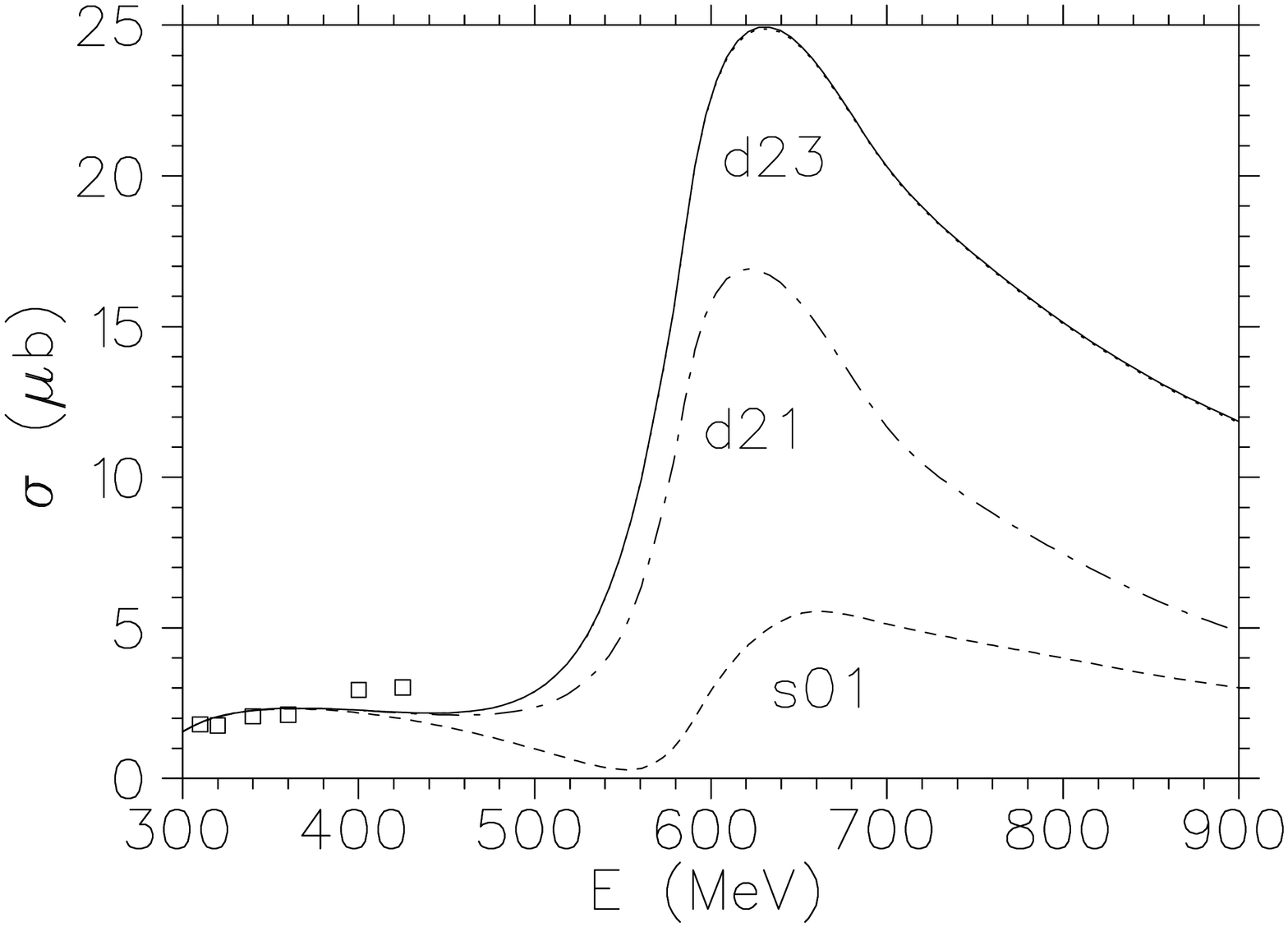}
\caption{The cumulative sum of the contributions to the overall
cross section from different partial wave amplitudes: solid
$s_{01}$, dashed $d_{21}$, dotted $d_{23}$. The solid curve denoting
the total integrated cross section including also the $g$-wave pions
is indistinguishable from the dotted one. The data are obtained by
integrating the fits of Ref. \cite{Bilger}.} \label{fig:partials}
\end{center}
\end{figure}

With such a strong variation and interference it is also
relevant to divide the cross section into partial waves to find
the important ones. As can be seen in Fig. \ref{fig:partials}
all partial wave amplitudes
 $s_{01},\; d_{21} \; {\rm and}\; d_{23}$ (in the notation
 ${l_\pi}_{JL}$) are about equally important at 800 MeV. In this figure
 the contribution of each partial wave to the angle integrated cross
section is presented as a function of the incident laboratory
energy. The $g$-wave pions contribute negligibly. The low-energy
Celsius data \cite{Bilger} are reasonably well reproduced.
Unfortunately there are no comparable data at other energies.

Further, a drastic energy dependence between 500 and 800 MeV is
found. This may be related to the constraint of small energy of the
final state diproton. Namely in that case the phase space integral
is very limited so that the angle integrated cross section is nearly
just a sum of the squared reaction matrix elements taken with the
maximal pion momentum. In this way the role of the $d$-wave pions is
emphasized. Also the contribution of a single partial wave can reach
even zero as seems to happen for the $s_{01}$ pions. In an
incoherent sum this could not happen. Similarly a phase space
integral over a wider range of momenta would probably have a
moderating effect in sharp changes and oscillations.

For orientation of the most imminent further experiments at COSY
\cite{Wilkin}, Fig. \ref{fig:forward} presents the energy dependence
of the forward cross section and its slope. The theoretical slope is
defined as the difference of the cross section at $\cos^2\theta =
0.9$ and $\cos^2\theta = 1$ divided by 0.1 (i.e. approximately the
derivative with respect to $\cos^2\theta$ but plotted as positive).
For the COSY data \cite{Dymov} the extremes of $\cos^2\theta$ were
used and for the CELSIUS the fits published in Ref. \cite{Bilger}.
If possible, the energy dependence is even more dramatic in this
quantity just in the experimentally uncharted region.

Here actual interference of all different partial waves is possible
and apparently at about 550 and 700 MeV the forward amplitude
changes its sign producing the small minima. Also, because the
forward cross section may be an order of magnitude smaller than the
bulk of the cross section as in Fig. \ref{fig:800mev}, due to
destructive interferences the absolute detailed prediction may not
be exactly correct, but still violent energy and angular variations
are expected. Certainly the expected behaviour of the cross section
is sharper than that in the widely studied $pp\rightarrow d\pi^+$
both as a function of energy and angle.

\section{Conclusion}
In summary, a phenomenological model calculation is performed for
the reaction $pp\rightarrow (pp)_{S{\rm -wave}}\pi^0$. Partly the
aim has been to provide some predictions in anticipation of
experiments at COSY. However, the finding of extreme energy and
angular dependencies may have also wider interest and applications
in other similar reactions in attempts to extract information on
reaction matrix elements. The constraint of a small relative
momentum for the final state protons seems to favour this strong
variation of the cross section in particular in the forward
direction but also in the angle-integrated cross section. Obviously
this cut also tends to stress higher pion waves than cross sections
integrated over all possible momenta. Such a strict constraint may
be a way to get hands on the squared reaction matrix elements at
(nearly) a single momentum choice.

\begin{figure}[t]
\begin{center}
\includegraphics[width=10 cm,clip= ]{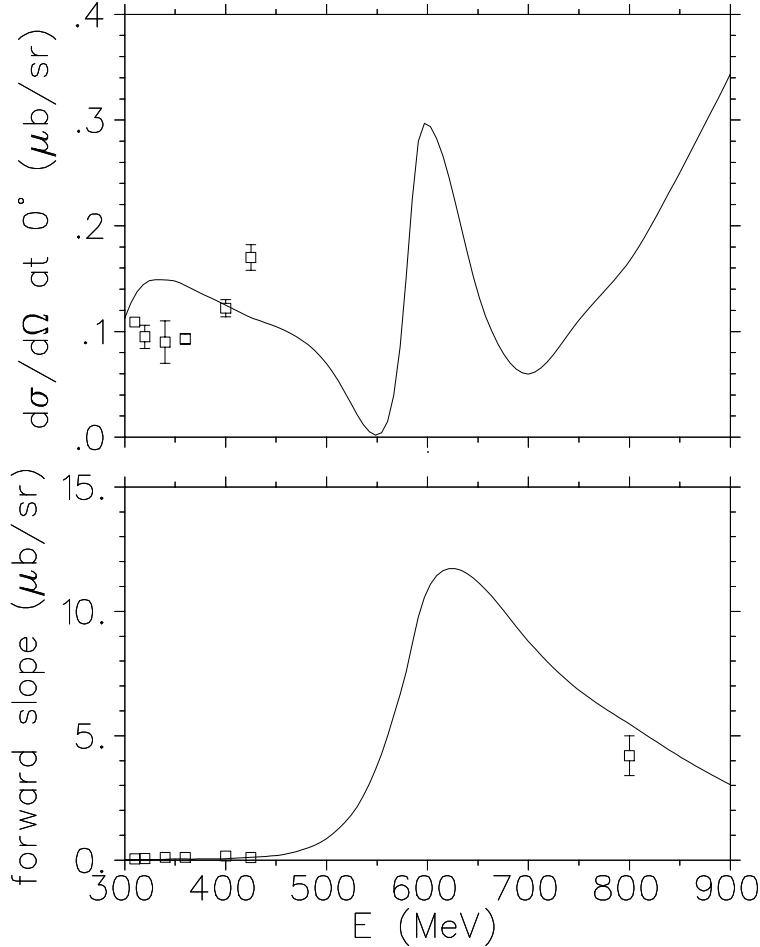}
\caption{Predictions for the forward cross section and its slope
defined as in the text. The data are from the fits of Ref.
\cite{Bilger} and from Ref. \cite{Dymov} (800 MeV). The forward
cross section at 800 MeV would be outside the figure at 700 nb/sr.}
\label{fig:forward}
\end{center}
\end{figure}

Already the experimental finding of opposite slopes of $d\sigma
/d\Omega$ with and without the cut on the final $pp$ excitation
energy is suggestive of physical differences. Calculations of total
production cross sections (integrated over all phase space) show
that already at 400 MeV the amplitudes
 $^1D_2\to {^3}P_2 s$, $^3P_1\to {^3}P_0 p$ and $^3F_3\to {^3}P_2 p$ are
 each as large as $^3P_0\to {^1}S_0 p$ with also sizable contributions
 from $^3P_2\to {^3}P_1 p$, $^3P_2\to {^3}P_2 p$ and $^3F_2\to {^3}P_2 p$.
  The $^1S_0s$ part of the cross section would then be
 only about one sixth of the total in line with the phenomenological fits
 of Ref. \cite{Bilger}. Apparently this complexity makes a detailed partial
 wave analysis improbable. However, with a cut, due to the additional
 simplicity of the spin structure of the final $^1S_0$ state, such
 an analysis is amenable with measurements of also spin observables.
 This, in turn, would act as a strong constraint on any modelling of
 the reaction over the whole phase space range.

The discrepancy with the COSY data in the forward direction may be
due to overly delicate destructive interference, which a minor
change in just one of the amplitudes might moderate. By some
exploratory model variations it was not possible to significantly
improve the situation. Making pion $s$-wave rescattering somewhat
stronger actually decreased the forward cross section. Change of the
size of the cut-off within the experimental precision has no
significant effect. An intriguing possibility could be a need for
explicit pion $d$-wave rescattering possibly involving the
 $N(1520)\; {\frac 1 2}^-$  resonance.

It would certainly be interesting to extend experiments both to
larger angles to check whether the model gives the total
normalization reasonably and also to the unexplored energies
accessible at COSY, where extreme energy dependence shown in Figs.
\ref{fig:partials} and \ref{fig:forward} is predicted. Further
details with model dependence and spin observables will be studied
in a forthcoming paper \cite{future}.

%\section{Acknowledgments}
%\begin{acknowledgments}
\begin{ack}
I thank C. Hanhart, Y. Uzikov and in particular C. Wilkin for
numerous useful discussions and suggestions for this work and J.
Zlomanczuk for providing the data of Ref. \cite{Bilger}. This work
was supported by the DAAD and Academy of Finland exchange programme
projects DB000379 (Germany) and 211592 (Finland).
 I also thank the Magnus Ehrnrooth Foundation for partial support
 and IKP of Forschungszentrum J\"ulich for hospitality.
%\end{acknowledgments}
\end{ack}

%\begin{references}

%\end{references}

\end{document}